\documentclass[]{raa}            
\usepackage{graphicx,times,subfig}
\usepackage{natbib}
\usepackage{longtable,textcomp}

\providecommand{\bm}[1]{\mbox{\boldmath$#1$\unboldmath}}

\begin{document}

     \title{Radio observations of the first three-month $\it{Fermi}$-AGN at 4.8 GHz}

 \volnopage{ {\bf 2011} Vol.\ {\bf 11} No. {\bf XX}, 000--000}
   \setcounter{page}{1}

   \author{Xiang Liu
          \inst{1}
           \and
           Hua-Gang Song\inst{1,2}
           \and
           Jun Liu\inst{1}
           \and
           Zhen Ding\inst{1}
           \and
            Nicola Marchili\inst{3}
           \and
           Thomas P. Krichbaum\inst{3}
           \and
           Lars Fuhrmann\inst{3}
           \and
           Anton Zensus\inst{3}
           \and
           Tao An\inst{4}
    }

   \institute{Xinjiang Astronomical Observatory, Chinese Academy of Sciences,
             150 Science 1-Street, Urumqi 830011, P.R. China; {\it liux@xao.ac.cn}\\
        \and Graduate University of the Chinese Academy of Sciences, Beijing 100049, P.R. China\\
        \and Max-Plank-Institut f\"ur Radioastronomie, Auf dem H\"ugel 69, 53121 Bonn, Germany\\
        \and Shanghai Observatory, Chinese Academy of Sciences, Shanghai 200030, P.R. China\\
\vs \no
   {\small Received [year] [month] [day]; accepted [year] [month] [day] }
}

\abstract{Using the Urumqi 25\,m radio telescope, sources from the
first three-month $\it{Fermi}$-LAT detected AGN catalog with declination
$>0^\circ$ were observed in 2009 at 4.8 GHz. The radio flux density
appears to correlate with the $\gamma$-ray intensity. Intra-day
variability (IDV) observations were performed in March, April and
May in 2009 for selected 42 $\gamma$-ray bright blazars, and $\sim$60\% of 
them show evident flux variability at 4.8 GHz during the IDV
observations, the IDV detection rate is higher than that in previous flat-spectrum AGN samples. 
The IDV appears more often in the VLBI-core
dominant blazars, and the non-IDV blazars show relatively `steeper' spectral
indices than the IDV blazars. Pronounced inter-month
variability has been found in two BL Lac objects: J0112+2244 and
J0238+1636.}

 \maketitle

 \keywords{galaxies: active -- quasars: general -- radio continuum: galaxies --gamma-rays: observations}

   \authorrunning{X. Liu et al.}            
   \titlerunning{Radio observations of the first three-month
                   $\it{Fermi}$-AGN at 4.8 GHz }


%
%
\section{Introduction}           
\label{sect:intro}

In the 1990s the space $\gamma$-ray telescope EGRET (the Energetic Gamma Ray Experiment Telescope) identified 66
blazars during its mission (Hartman et al. 1999). $\it{Fermi}$ (the Fermi Gamma Ray Space Telescope spacecraft),
the successor of EGRET, launched in 2008, is now in all-sky survey
mission. $\it{Fermi}$ has a much higher sensitivity and pointing
accuracy than EGRET, and has already detected more AGN (Active Galactic Nuclei), see Abdo et
al. (2010). The first three-month observations of the
$\it{Fermi}$-LAT (Large Area Telescope on board the Fermi Gamma Ray Space Telescope spacecraft) detected 132 bright
$\gamma$-ray sources in which 104 are blazars (Abdo et al. 2009).
Some of the $\it{Fermi}$-LAT detected AGN exhibit $\gamma$-ray variability on
timescales of days.

Blazars are either flat-spectrum radio quasars or BL Lac
objects, and they are extremely variable at all observable
wavelengths on timescales ranging from less than an hour to many
years, both in total power and linear polarization. The apparent
motions of VLBI (Very long Baseline Interferometry) components along their jets are often highly
superluminal with brightness temperatures being close to the
inverse-Compton limit. Such violent behaviour in blazars is
attributed to relativistic jets oriented very close to the line of
sight (Rees 1966; Urry \& Padovani 1995). The relationship of
variability at different wavelengths is a crucial test for
theoretical models of these outbursts/flares in $\gamma$-ray AGN.

Although it is generally accepted that the $\gamma$-rays detected
from blazars are emitted from collimated jets of charged particles
moving at relativistic speeds (Maraschi et al. 1992), open
questions remain. The mechanisms by which the particles are
accelerated, the precise site of the $\gamma$-ray production, the
origin of AGN variability and the $\gamma$-ray duty cycle of blazars
are still not well understood.

IDV (Intra-Day Variability, rapid variability on timescales of
few hours to few days) observation and radio monitoring can provide
the variability characteristics of AGN in radio, allowing us to search for
correlations between radio and $\gamma$-ray luminosities and to study
the connection between the emission mechanisms.

Intraday variability of radio flux density has been found in
about 30\% to 50\% of the flat-spectrum radio sources (Quirrenbach
et al. 1992; Lovell et al. 2008). If interpreted as being source
intrinsic, the rapid variability would imply micro-arcsecond scale
sizes of the emitting regions, which would result in excessively
large apparent brightness temperatures far in excess of the
inverse-Compton limit of $\sim\rm10^{12} K$ (Kellermann \&
Pauliny-Toth 1969; Readhead 1994). Thus, theories which explain
IDV with variations intrinsic to the blazars, require either
excessively large Doppler boosting factors or special source
geometries (such as non-spherical relativistic emission models,
e.g. Qian et al. 1996) or coherent and collective plasma emission
(Benford 1992; Lesch \& Pohl 1992) to avoid the inverse-Compton
catastrophe. Alternatively, IDV was explained by interstellar
scintillation (ISS), especially for very rapid variables such as
PKS 0405$-$385, J1819+384, PKS 1257$-$326 and J1128+592
(Kedziora-Chudczer et al. 1997; Dennett-Thorpe \& de Bruyn 2000;
Bignall et al. 2003; Gabanyi et al. 2009 respectively).

Almost all the $\it{Fermi}$-LAT detected AGN are blazars. There
appears a significant correlation between the radio flux density at 15
GHz and $\gamma$-ray flux density of the $\it{Fermi}$-LAT AGN
(Ackermann et al. 2011). It is expected that the
$\it{Fermi}$-AGN are more IDV-active than non $\gamma$-ray AGN.
Dedicated IDV and flux density monitoring observations are needed
to study the variability on different timescales and to correlate
the occurrence of IDV with the $\gamma$-ray activity of the
$\it{Fermi}$-AGN. In March 2009, we launched a program with the
25-m Urumqi radio telescope at 4.8 GHz to investigate the intra-day
to inter-month variability of the first three-month
$\it{Fermi}$-detected AGN. We aim at searching for new IDV
sources, and for a statistical comparison of the radio and the
$\gamma$-ray emission of $\it{Fermi}$-detected AGN.

In this paper, we present the results from our single dish radio
observations of the $\it{Fermi}$-LAT detected AGN with declination $>0^\circ$.
We adopt $S \propto \nu^{\alpha}$ to define a spectral index
throughout the paper.


\section{The sample and observations}
\label{sect:Obs}

Our sample is selected from the first three-month $\it{Fermi}$-LAT detected AGN
catalog (Abdo et al. 2009). We originally observed 63 sources
with declination $>0^\circ$ from the catalog as our pilot `cross-scan' observations in
March 2009. Thirteen sources were rejected due to poor data quality in the pilot run.
Flux densities of 50 sources were obtained in the pilot observation at
4.8 GHz, including the galaxy NGC1275 (3C84). We selected the sources 
for IDV observation from the 50 sources by following criteria:

\begin{enumerate}
 \item
Blazars with source brightness: $S_{4.8 GHz} >$ 0.3 Jy.
\item
Source compactness: measured full-width-half-maximum (FWHM) of source brightness profile $<$ 700\textquotesingle\textquotesingle\, 
 which is $\sim$1.2 times the antenna beam size at 4.8 GHz. 
Extended source brightness profile may indicate a confusion
between the target and its nearby sources which cannot be resolved by the 25-m radio telescope at 4.8 GHz.
\end{enumerate}

The criteria restrict the number of the sources for IDV observation to be 45, unfortunately 3 of them, namely J0920+4441, J1229+0203 (3C273) and 
J1253+5301 were not involved in afterwards IDV campaign by mistake. Therefore, the sources which IDV observations 
were carried out are 42, which consists of 24
flat-spectrum radio quasars and 18 BL Lac objects. 

The IDV observations were carried out in
order to study the short time-scale variability of the
$\gamma$-ray bright blazars. These sources were also planned to
monitor monthly from March to December in 2009 at 4.8 GHz.
Since all of the selected sources are strong and compact, both the
IDV observations and the flux monitoring were performed in
`cross-scan' mode.

\subsection{IDV observations}

The three IDV observing sessions performed with the Urumqi
telescope are summarized in Table~\ref{table:idvobs}; column 1
the symbols for different epochs; column 2 the starting and
ending dates of the experiments; column 3 the duration; column 4
the mean number of flux density measurements per hour; column 5
the number of observed sources (including the calibrators, usually
we use 3C286, 3C48 and NGC7027 as primary calibrators, B0836+710
and B0951+699 etc as secondary calibrators); column 6 the average
number of measurements per hour for each $\it Fermi$ blazar (duty
cycle, which represents the shortest time scale on which we can
search for variability); column 7 the average modulation index of calibrators, which is
probably the most important since it reflects the conditions of
the observation (the lower the $m_0$ the better the weather,
and/or the more stable the receiver) and will be described in the
following sections.
\begin{center}
\begin{table}
\caption{Summary of the IDV observations for the 42 selected $\it
Fermi$ blazars} \label{table:idvobs}
\begin{tabular}{ccccccc}\hline \noalign{\smallskip}
Epoch & Date & Duration & Average & Number of & Duty cycle for & $m_0$\\
 &  & [$d$] & sampling [$h^{-1}$] & observed sources & $\it Fermi$ blazars [$h^{-1}$] & [\%]\\
 (1)&(2) &(3) &(4) & (5)&(6) &(7) \\
\hline \noalign{\smallskip}
A & 21/03/09 - 25/03/09 & 4.8 & 11.2 & 49 & 0.3 & 0.6 \\
B & 19/04/09 - 24/04/09 & 5.3 & 10.8 & 37 & 0.4 & 0.7 \\
C & 06/05/09 - 09/05/09 & 3.8 & 10.1 & 17 & 0.8 & 0.6 \\
\hline \noalign{\smallskip}
\end{tabular}
\end{table}
\end{center}

\subsection{Inter-month observations}

The 42-blazar sample was also planned to be monitored monthly from
March to December in 2009 at 4.8 GHz. Each source was at least
measured once in an individual observation. Sometimes the flux
density measurements were repeated for sources whose observations
were affected by adverse conditions such as bad weather and low
elevation. Actually, not all sources were observed
every month due to time limitation, so that some
sources have no data in some months. In fact no data in June 2009
for all sources.

\section{Data calibration}


All observations have been done in `cross-scan' mode with a central frequency of 
4800 MHz and a bandwidth of 600 MHz, see Sun et al. (2007) for a description of the 
observing system. Each
scan consists of 8 sub-scans in azimuth and elevation over the
source position, fourfold in each coordinate.
This enables us to check the pointing offsets in both coordinates.
After applying a correction for small pointing offsets, the
amplitudes of both azimuth and elevation are averaged. Then, we
correct the measurements for the elevation dependent antenna gain
and the remaining systematic time-dependent effects by using a
number of steep spectrum and non-variable secondary calibrators.
Finally, we convert our measurements to absolute flux density. The
conversion factor is determined as the average scale of the
frequently observed primary calibrator's assumed flux densities
(Baars et al. 1977; Ott et al. 1994) by the measured temperatures, 
where we use the assumed flux density of 7.53 Jy, 5.53 Jy and 5.47 Jy 
at 4.8 GHz for the primary calibrators 3C286, 3C48 and NGC7027 respectively in our data reduction.

The overall typical error on a single measurement is around
0.3-1.5\% of source flux density depending e.g. on weather
conditions and source intensity, usually weaker sources have
larger errors in individual measurements.
Our data reduction of the radio flux density has an essentially
similar procedure to the Effelsberg data reduction (e.g. Kraus et
al. 2003). Actually we have made simultaneous IDV observations with
Urumqi and Effelsberg telescopes as early as April 2006, we obtained consistent results
between the two telescopes.  

\begin{figure}[ht!]
\centering
     \includegraphics[width=12cm]{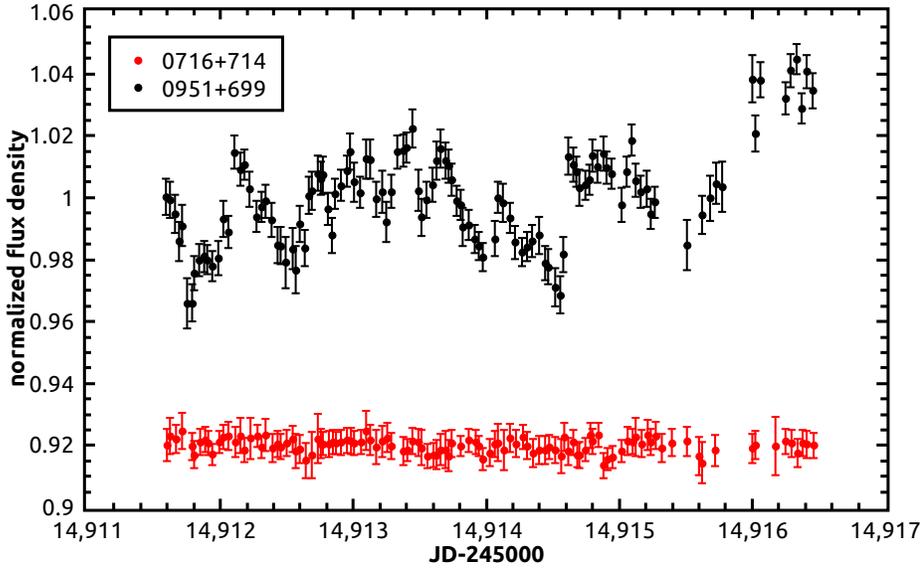}
     \caption{The variability curve (normalized flux density $S/\langle\,S\,\rangle$ versus time) of 
J0721+7120 (B0716+714, upper),
       together with a nearby secondary calibrator B0951+699 (lower) in epoch A. The offset of the
       two curves is arbitrary, for better visualization.}
      \label{fig:fig1}
   \end{figure}

After the data reduction described above, the light curves of the
sources were obtained. In Fig.~\ref{fig:fig1} we give an
example of the variability curve after data calibration, we can
see that the scatter of the calibrator is really small.

In addition, 11 sources have been measured for flux densities while they were not involved in the IDV observations, 
the flux densities are useful in the correlation analysis between radio flux density and $\it{Fermi}$ $\gamma$-ray intensity. 
So we list the source name/type (Q-quasar, BL-BL Lac object), flux density and observing epoch in Table~\ref{table:add10}, 
although three sources were measured for flux densities in July, August and September 2009. 
The radio galaxy 3C84 has the flux density of 16.174$\pm$0.065 Jy measured in March 9, 2009, which 
is not listed in Table~\ref{table:add10}.

\begin{center}
\begin{table}
\caption{Flux densities measured for 10 blazars while they were not involved in IDV observations} \label{table:add10}
\begin{tabular}{cccccc}
\hline \noalign{\smallskip}
\hline \noalign{\smallskip}
Name/type  &J0144+2705/BL &J0654+4514/Q &J0920+4441/Q &J0948+0022/Q &J1058+5628/BL \\
S(Jy) & 0.240$\pm$0.003 & 0.449$\pm$0.003 &1.058$\pm$0.006  &0.191$\pm$0.006  & 0.189$\pm$0.007  \\
Epoch & 20090309 & 20090723 & 20090311 & 20090311 &  20090311  \\
\hline \noalign{\smallskip}
Name/type  &J1229+0203/Q &J1253+5301/BL &J1542+6129/BL  &J1959+6508/BL &J2325+3957/BL \\
S(Jy) & 40.174$\pm$0.165 & 0.345$\pm$0.009 & 0.142$\pm$0.005 &0.228$\pm$0.004 &0.200$\pm$0.005   \\
Epoch & 20090311 & 20090311 & 20090919 & 20090309 &  20090819  \\
\hline \noalign{\smallskip}
\end{tabular}
\end{table}
\end{center}

\section{Statistical analysis and results}

For the data analysis, we use several quantities such as the
modulation index, the variability amplitude and so forth. Here we
just give a brief summary, a detailed description of these
parameters can be found in Kraus et al. (2003).
\begin{enumerate}
 \item The modulation index $m$

$m$ is defined as the standard deviation of the flux density divided by
the mean flux density of source.

\begin{displaymath}
m[\%]=100\cdot\frac{\sigma_S}{\langle\,S\,\rangle},
\end{displaymath}
$m$ provides a measure of the
strength of the variations observed. One should notice that it
does not account for the intrinsic noise in the data.

 \item The variability amplitude $Y$

The noise-bias corrected variability amplitude $Y$ is defined as
\begin{displaymath}
Y[\%]=3\sqrt{m^{2}-m_0^2},
\end{displaymath}
where $m_0$ is the mean modulation index of all non-variable
sources (see Heeschen et al. 1987), so, it corresponds to the
average residual noise and possibly even systematic scatter in the
calibrators. Thus, compared to $m_0$, $Y$ works as a more uniform
estimator of the variability strength, and it is useful
for comparing data of different epochs.

\item $\chi^2$ and reduced $\chi^2$

As a criterion for the source variability, the hypothesis of a constant
function is examined and the calibrated data is fitted by a $\chi^2$ test
of the kind
\begin{displaymath}
\chi^2=\sum(\frac{S_i-\langle\,S\,\rangle}{\Delta\,S_i})^2,
\end{displaymath}
with the reduced $\chi^2$
\begin{displaymath}
\chi_r^2=\frac{1}{N-1}\cdot\sum(\frac{S_i-\langle\,S\,\rangle}{\Delta\,S_i})^2,
\end{displaymath}
where $S_i$ denotes the individual flux densities, $\langle\,S\,\rangle$ their
average in time, $\Delta\,S_i$ their errors and N the number of
measurements (e.g. Bevington \& Robertson 1992). Only those
sources for which the probability that they can be fitted by a
constant function is $\leq$\,0.1\%, are considered to be variable.
\end{enumerate}

We define the modulation index $M$ for the inter-month flux
density variations similar as $m$ for the IDV. For the calibrators
we obtain $M_0\sim$2 [\%]. Because the inter-month data are sparse
and unevenly sampled, we just treat the $M_0$ as the error of $M$,
and did not define a noise-bias corrected $Y$ for the inter-month
data.

Here, we present the basic information of the sources, and the
statistical results of the IDV and inter-month observations at 4.8
GHz in Table~\ref{table:flux}. The different columns are assigned
as follows: column (1) source J2000 name; (2) optical
identification (Q: quasar, BL: BL Lac object) and VLBI structure 
(c: extremely core dominated, cj: core with a mild jet); 
(3) the spectral index from the SPECFIND V2.0 catalog of
broad band radio spectra (Vollmer et al. 2010), by using a least squares fit of the broad band 
[70-10500] MHz data for the majority of sources, and that the spectral index
 from two frequencies 1.4 and 8.4 GHz which is from the CRATES catalog (Combined
Radio All-Sky Targeted Eight GHz Survey, Healey et al. 2007) except for three
 sources where the information is from NED; (4) symbols for different epochs (see
Table~\ref{table:idvobs}); (5) effective number of flux measurements in
IDV observation; (6) mean flux density in IDV observation; (7)
the modulation index of IDV; (8) the reduced $\chi^2$ of IDV; (9)
the variability amplitude of IDV, a zero is given if $m \leq m_0$;
(10) IDV identification, a `$+$' is given if the source shows IDV;
(11) the mean flux density of the inter-month observations and the
modulation index of inter-month variability; 
(12) the effective number of measurements for inter-month variability.

\clearpage
\begin{small}\begin{center}
\begin{longtable}[]{ccc|ccccccc|cc}
\caption{The results of IDV and inter-month radio observations}
\label{table:flux}
\endfirsthead
\multicolumn{12}{c}
{{\bfseries \tablename\ \thetable{}
: The results of IDV and inter-month radio observations -- continued}} \\

\hline\hline\noalign{\smallskip} Source & Id/vlbi& $\alpha1^{0}/\alpha2$ & Ep &
N & $\overline{S}$[Jy] & $m$[\%] & $\chi_{r}^{2}$
&$Y$[\%] & IDV & $\overline{S_{M}}$[Jy]/$M$[\%]& No.\\
(1)&(2) &(3) &(4) & (5)&(6) &(7) & (8)&(9) &(10) &(11) &(12)\\
\hline\noalign{\smallskip}
\endhead
\hline\noalign{\smallskip}
\endfoot
\hline\noalign{\smallskip}
\endlastfoot
\hline\hline\noalign{\smallskip} Source & Id/vlbi& $\alpha1^{0}/\alpha2$ & Ep &
N & $\overline{S}$[Jy] & $m$[\%] & $\chi_{r}^{2}$
&$Y$[\%] & IDV & $\overline{S_{M}}$[Jy]/$M$[\%]& No.\\
(1)&(2) &(3) &(4) & (5)&(6) &(7) & (8)&(9) &(10) &(11) &(12)\\
\hline\noalign{\smallskip}

  J0112+2244   & BL/c &  0.18$^{a}$/0.12   &                A  &  10   &  0.756  &   3.02 & 16.42  &  8.89  &  $+$  & 0.508/37.1&7\\
   &&&                                        B  &  10   &  0.674  &   6.21 & 30.56  & 18.51  &  $+$  & \\
  J0136+4751  & Q/c  &  0.10/0.19   &                A  &  13   &  4.320  &   1.18 &  9.93  &  3.04  &  $+$  & 4.044/6.6&8\\
  J0217+0144   & Q/c  &  0.16/0.24   &                A  &  7    &  1.298  &   2.17 & 12.74  &  6.24  &  $+$  & 1.284/7.0&7\\
  &&&                                        B  &  14   &  1.288  &   3.75 & 23.88  & 11.06  &  $+$  &  \\
  J0238+1636  & BL/c  &  0.26/0.56   &                A  &  10   &  3.331  &   1.00 &  7.03  &  2.40  &  $+$  & 2.225/42.7&7\\
  J0530+1331  & Q/cj   &  0.44$^{a}$/0.24   &                A  &   9   &  3.490  &   0.34 &  0.75  &  0.00  &       & 3.121/9.9&8\\
  &&&                                        B  &  15   &  3.587  &   2.95 & 34.38  &  8.61  &  $+$  & \\
  J0654+5042  & Q/cj   & 0.30$^{a}$/0.23   &                A  &   9   &  0.315  &   1.57 &  1.23  &  4.37  &       & 0.313/10.2&7\\
  &&&                                        B  &  20   &  0.303  &   2.76 &  2.98  &  8.01  &  $+$  & \\
  J0712+5033  & BL/c  &  0.37$^{a}$/0.40   &                A  &  12   &  0.313  &   2.79 &  4.13  &  8.17  &  $+$  & 0.315/9.4&2\\
  &&&                                        B  &  25   &  0.349  &   3.39 &  5.01  &  9.94  &  $+$  & \\
  J0713+1935  & Q/c   &  0.16$^{b}$/0.35  &                 A  &   9   &  0.276  &   6.81 & 17.04  & 20.34  &  $+$  & 0.254/19.1&2\\
  &&&                                        B  &  15   &  0.451  &   6.99 & 34.75  & 20.87  &  $+$  & \\
  J0719+3307  & Q/c   & -0.20/-0.15   &                A  &  10   &  0.509  &   2.17 &  4.98  &  6.26  &  $+$  & 0.546/6.9&2\\
  &&&                                        B  &  24   &  0.595  &   3.83 & 13.53  & 11.30  &  $+$  & \\
  J0721+7120  & BL/c  & -0.43$^{c}$/-0.13  &                A  & 117   &  1.241  &   1.61 &  8.34  &  4.48  &  $+$  & 1.305/18.4&9\\
  &&&                                        B  &  92   &  1.361  &   3.66 & 37.68  & 10.77  &  $+$  & \\
  &&&                                        C  &  85   &  1.134  &   3.35 & 25.11  &  9.90  &  $+$  & \\
  J0738+1742  & BL/cj  & -0.01/0.27   &                A  &   8   &  0.902  &   0.54 &  0.68  &  0.00  &       & 0.915/2.0&2\\
  &&&                                        B  &  16   &  0.935  &   2.75 & 14.26  &  7.97  &  $+$  & \\
  J0818+4222  & BL/cj  & -0.10/-0.04   &                A  &  10   &  1.503  &   1.90 & 15.73  &  5.40  &  $+$  & 1.660/11.9&8\\
  &&&                                        B  &  24   &  1.522  &   2.03 &  8.07  &  5.71  &  $+$  & \\
  J0824+5552  & Q/cj   & -0.08/0.10   &                A  &  13   &  1.049  &   0.59 &  0.92  &  0.00  &       & 1.083/6.7&7\\
  &&&                                        B  &  28   &  1.048  &   0.66 &  0.66  &  0.00  &       & \\
  J0854+2006   & BL/cj & 0.35/0.44   &                A  &   8   &  1.952  &   1.56 & 12.16  &  4.31  &  $+$  & 1.911/5.1&2\\
  J0957+5522  & Q/cj   & -0.34$^{c}$/-0.41   &                A  &  13   &  1.939  &   0.15 &  0.08  &  0.00  &       & 1.980/4.6&8\\
  &&&                                        B  &  26   &  1.937  &   0.30 &  0.28  &  0.00  &       & \\
  J1015+4926  & BL/cj  & -0.21/-0.24   &                A  &  10   &  0.358  &   1.48 &  1.15  &  4.04  &       & 0.352/4.0&4\\
  &&&                                        B  &  19   &  0.356  &   1.50 &  1.06  &  3.99  &       & \\
  J1016+0513  & Q/cj   & -0.07$^{a}$/-0.18   &                A  &   4   &  0.533  &   1.43 &  2.56  &  3.90  &       & 0.593/5.9&3\\
  J1033+6051  & Q/cj   & -0.21$^{c}$/-0.05   &                A  &  17   &  0.427  &   2.99 &  5.22  &  8.77  &  $+$  & 0.382/16.7&7\\
  &&&                                        B  &  34   &  0.412  &   2.89 &  4.61  &  8.41  &  $+$  & \\
  J1104+3812  & BL/cj  & -0.25/-0.11   &                A  &   8   &  0.590  &   1.03 &  1.36  &  2.51  &       & 0.596/3.0&5\\
  J1159+2914  & Q/c   & -0.29/-0.29   &                A  &  62   &  2.733  &   1.86 & 14.31  &  5.29  &  $+$  & 2.702/2.6&2\\
  &&&                                        B  &  55   &  2.627  &   1.30 &  4.69  &  3.29  &  $+$  & \\
  J1217+3007  & BL/cj  & -0.33/-0.30   &                A  &  10   &  0.443  &   1.41 &  1.39  &  3.82  &       & 0.452/7.8&5 \\
  &&&                                        B  &  19   &  0.441  &   1.55 &  1.54  &  4.13  &       & \\
  J1221+2813  & BL/cj  &  0.07/0.19   &                A  &  11   &  0.481  &   1.82 &  2.58  &  5.17  &       & 0.491/2.0&2\\
  J1310+3220  & Q/cj   &  0.09/0.30   &                A  &   9   &  1.054  &   2.00 &  7.06  &  5.73  &  $+$  & 1.222/8.7&6\\
  &&&                                        B  &  18   &  1.092  &   1.24 &  2.73  &  3.06  &  $+$  & \\
  J1427+2348  & BL/cj  & -0.33/-0.34   &                A  &  10   &  0.362  &   2.20 &  2.38  &  6.36  &       & 0.340/1.5&3\\
  &&&                                        B  &  19   &  0.355  &   1.83 &  1.47  &  5.07  &       & \\
  &&&                                        C  &   2   &  0.359  &   1.16 &  0.27  &  2.96  &       & \\
  J1504+1029  & Q/cj   & 0.06/-0.03    &                A  &   8   &  1.524  &   0.63 &  0.77  &  0.54  &       & 1.601/8.2&9\\
  &&&                                        B  &  14   &  1.611  &   0.86 &  1.58  &  1.50  &       &  \\
  &&&                                        C  &  31   &  1.725  &   0.94 &  1.98  &  2.17  &       &  \\
  J1522+3144  & Q/c   & -0.14/0.18   &                A  &   9   &  0.536  &   0.55 &  0.39  &  0.00  &       & 0.528/7.9&8\\
  &&&                                        C  &  33   &  0.512  &   2.60 &  4.61  &  7.60  &  $+$  & \\
  J1553+1256  & Q/cj   & -0.26/-0.47   &                A  &   8   &  0.703  &   0.48 &  0.10  &  0.00  &       & 0.723/9.3&8\\
  &&&                                        C  &  29   &  0.721  &   1.37 &  2.40  &  3.71  &  $+$  & \\
  J1555+1111  & BL/c  &  0.04/0.26   &                C  &  13   &  0.315  &   2.79 &  1.46  &  8.18  &       & 0.298/4.8&4\\
  J1635+3808  & Q/cj   &  -0.01/-0.09   &                A  &  11   &  2.894  &   0.33 &  0.53  &  0.00  &       & 3.165/10.6&8\\
  &&&                                        B  &  21   &  2.931  &   0.38 &  0.55  &  0.00  &       & \\
  &&&                                        C  &  33   &  3.013  &   0.45 &  0.76  &  0.00  &       & \\
  J1653+3945  & BL/cj  & -0.12/-0.19   &                A  &  10   &  1.535  &   0.58 &  0.95  &  0.00  &       & 1.573/7.7&8 \\
  &&&                                        C  &  32   &  1.539  &   0.60 &  0.73  &  0.00  &       & \\
  J1719+1745  & BL/cj  &  0.21$^{a}$/0.03   &                A  &  10   &  0.592  &   2.64 &  5.12  &  7.72  &  $+$  & 0.621/8.5&7\\
  &&&                                        B  &  19   &  0.590  &   1.77 &  3.16  &  4.87  &  $+$  & \\
  &&&                                        C  &  27   &  0.603  &   1.48 &  2.11  &  4.06  &  $+$  & \\
  J1751+0939  & BL/c  &  0.41$^{a}$/0.64   &                A  & 10    &  2.872  &   1.37 &  9.51  &  3.69  &  $+$  & 3.204/16.8&3\\
  &&&                                        C  & 27    &  2.572  &   1.36 &  7.32  &  3.67  &  $+$  & \\
  J1800+7828  & BL/cj  &  0.07$^{c}$/0.13   &                A  & 23    &  2.233  &   0.42 &  0.74  &  0.00  &       & 2.208/4.3&3\\
  &&&                                        B  & 42    &  2.128  &   0.50 &  0.68  &  0.00  &       & \\
  &&&                                        C  & 83    &  2.183  &   0.72 &  1.34  &  1.21  &       & \\
  J1848+3219  & Q/cj   & -0.16$^{a}$/0.11   &                A  &  9    &  0.557  &   1.73 &  2.83  &  4.87  &       & 0.619/8.2&3\\
  &&&                                        B  & 21    &  0.600  &   1.41 &  2.31  &  3.66  &  $+$  & \\
  &&&                                        C  & 30    &  0.623  &   1.15 &  1.48  &  2.93  &       & \\
  J1849+6705  & Q/c   & -0.20$^{c}$/-0.06   &                A  & 22    &  1.240  &   0.83 &  2.07  &  1.71  &       & 1.280/7.4&8\\
  &&&                                        B  & 40    &  1.272  &   1.79 &  7.69  &  4.95  &  $+$  & \\
  &&&                                        C  & 77    &  1.282  &   1.90 &  9.12  &  5.41  &  $+$  & \\
  J2147+0929  & Q/c   & -0.09/0.03   &                A  & 11    &  0.782  &   2.25 &  7.83  &  6.49  &  $+$  & 0.983/16.3&5\\
  &&&                                        B  & 15    &  0.891  &   2.96 & 13.61  &  8.64  &  $+$  & \\
  J2157+3127  &  Q/c  & -0.16/0.07   &                A  & 7     &  0.399  &   1.19 &  1.03  &  3.09  &       & 0.480/17.5&7\\
  J2202+4216  & BL/cj  &  0.17/0.31   &                A  & 10    &  2.950  &   1.16 & 15.04  &  2.99  &  $+$  & 3.506/21.8&3\\
  &&&                                        B  & 24    &  3.125  &   0.57 &  1.25  &  0.00  &       & \\
  J2203+1725  & Q/c   &  0.00/0.32   &                A  & 12    &  1.022  &   2.15 & 14.05  &  6.21  &  $+$  & 0.972/5.0&2\\
  &&&                                        B  & 18    &  0.953  &   2.27 &  9.99  &  6.48  &  $+$  & \\
  J2232+1143  & Q/cj  & -0.14/-0.42   &                A  &  7    &  4.605  &   0.53 &  1.48  &  0.00  &       &  4.794/2.4&2\\
  &&&                                        B  & 17    &  4.663  &   0.35 &  0.44  &  0.00  &       &  \\
  J2253+1608  & Q/cj   & -0.04/-0.11   &                A  & 12    & 10.417  &   0.54 &  2.11  &  0.00  &       & 10.390/6.1&7\\
  J2327+0940  & Q/cj   & -0.03$^{a}$/-0.08   &                A  &  7    & 1.372   &   9.35 &125.19  & 27.98  &  $+$  &   1.379/2.5&2\\
  &&&                                        B  & 18    & 1.419   &   1.02 &  2.35  &  2.23  &       &   \\

\end{longtable}
\end{center}            \end{small}
\footnotetext{a--frequency range [300-10500] MHz, 
b--frequency range [1000-10000] MHz, c--frequency range [30-10500] MHz, 
others have frequency ranges of [70-10500] MHz, for $\alpha1$ in the column 3 of Table~\ref{table:flux}. }

As shown in Table~\ref{table:flux}, 26 sources (16 QSOs and 10 BL
Lacs) show intra-day variability at a confidence level of larger
than 99.9\% at least once in IDV observations, according to
$\chi^2$ test. Thus the IDV occurrence is 26 out of 42
$\it{Fermi}$-blazars, indicating the IDV detection rate is $\sim$60\%,
which is higher than that in previous flat-spectrum AGN samples
(e.g. Quirrenbach et al. 1992; Lovell et al. 2008). This high rate could be
caused by a higher compactness of $\it{Fermi}$ blazars relative to
sources in other samples.

We find very pronounced inter-month variability in two BL Lac
objects: J0112+2244 and J0238+1636, with modulation index 37.1\%
and 42.7\%, respectively. Because the inter-month observation data
are sparse and unevenly sampled, we will use only the 24 sources
which have been observed at least in 5 months, for statistics in
the following.

\subsection{$\gamma$-ray and radio flux density}

\begin{figure}[ht!]
\centering
     \includegraphics[width=10cm]{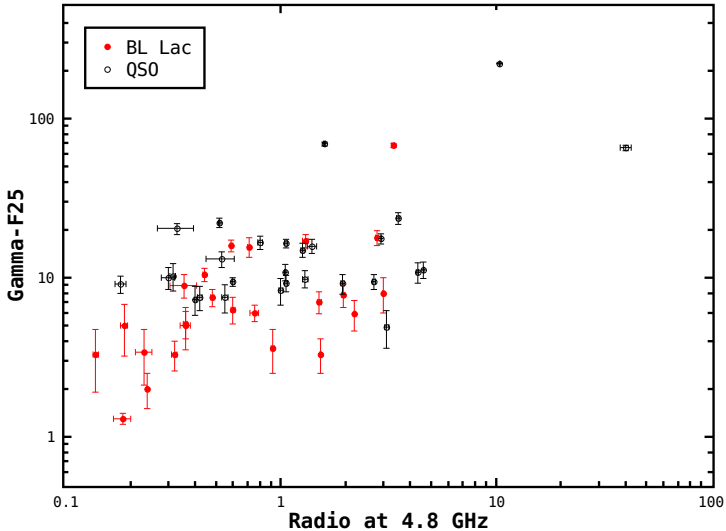}
     \caption{The mean flux densities (averaged over all IDV sessions) and those from Table~\ref{table:add10} at 4.8 GHz vs.
      the gamma-ray intensities ($\geq$100 MeV, in $10^{-8}ph\, cm^{-2}s^{-1}$, from Abdo et al. 2009) of 52
      $\it{Fermi}$-blazars; black: QSO, red: BL Lac.}
      \label{fig:fig2}
   \end{figure}

Combined radio and $\gamma$-ray data can be used to study the
relationship between radio and $\gamma$-ray emission. We find
that there is a correlation between the 4.8 GHz radio flux density
of 52 sources and their $\gamma$-ray intensity
(Fig.~\ref{fig:fig2}) with a Spearman correlation coefficient of
0.48 (significance 2.8$\times10^{-4}$) in total, 0.37
(significance 0.06) for QSOs, and 0.61 (significance 0.001) for BL
Lacs, respectively. For quasars the correlation seems not significant, especially when
removing the two strong quasars 3C273 and 3C454.3. 
However there still has a correlation coefficient
of 0.42 (significance 0.02) for 50 sources after dropping the 2
strong quasars. It is notable that our sample is incomplete and the radio-$\gamma$
data are not simultaneous. The 15 GHz simultaneous radio-$\gamma$
result obtained in Caltech group (Readhead, private communication)
shows a more significant correlation, where a new Monte-Carlo
method was applied and an intrinsic correlation was found
(also see Ackermann et al. 2011).

The correlation between the radio and the $\gamma$-ray
emission in the blazars can shed light on the
physical link between the emission processes in the two energy
bands. It is suggested that the $\it{Fermi}$-LAT detected blazars have on
average higher Doppler factors than non-$\it{Fermi}$-LAT detected blazars
(Savolainen et al. 2010). It is possible that the $\gamma$-ray
emission (via synchrotron self-Compton and/or inverse-Compton
scattering by the relativistic electrons in radio jet) from
the $\it{Fermi}$-LAT detected blazars could also be Doppler-boosted. 

\subsection{Spectral index and flux density variability}


It is possible that the 4.8 GHz flux density variability is
related to the source spectral index. In order to obtain a more
realistic estimate of the spectrum from broad-band radio data, we checked our 42 blazars in
SPECFIND V2.0 catalog of broad band [30-10500] MHz radio continuum
spectra (Vollmer et al. 2010), and obtained their indices, as
the $\alpha1$ shown in Table~\ref{table:flux}. The frequency ranges of the spectra are
[70-10500] MHz for 27 sources, [300-10500] MHz for 9 sources, 
[30-10500] MHz for 5 sources, and [1000-10000] MHz for 1 source, respectively.
Although the spectral indices
are from relatively low frequency ranges, the 42 blazars still
show flat spectral indices $\alpha1>-0.5$ from the SPECFIND V2.0
catalog. We also list the spectral index $\alpha2$ calculated with 1.4 GHz
and 8.4 GHz flux density from the CRATES catalog in Table~\ref{table:flux}.

\begin{figure}
\centering
\hspace{0pt}%
\subfloat{%
\includegraphics[width=0.48\textwidth]{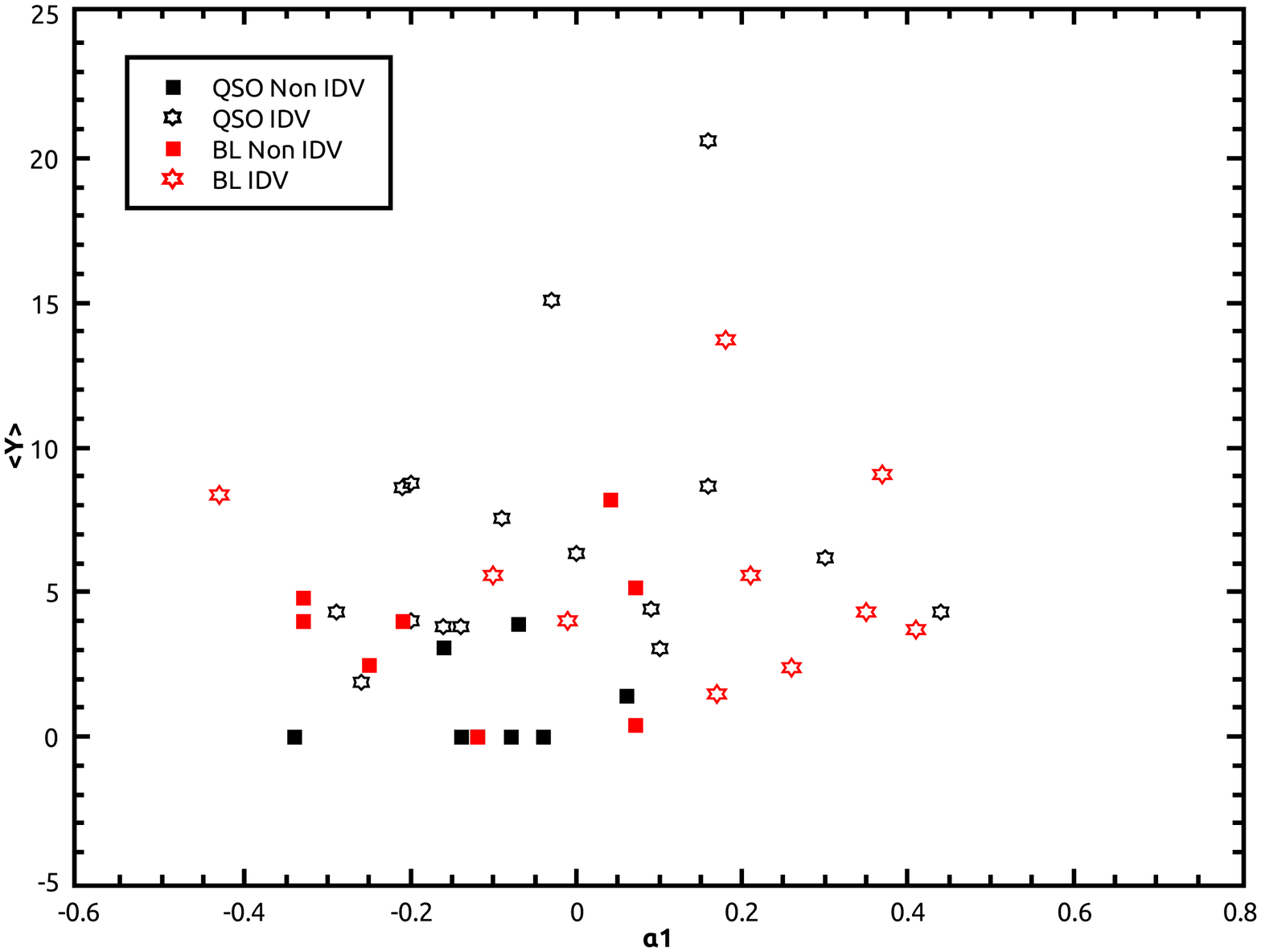}}
\hspace{0pt}%
\subfloat{%
\includegraphics[width=0.48\textwidth]{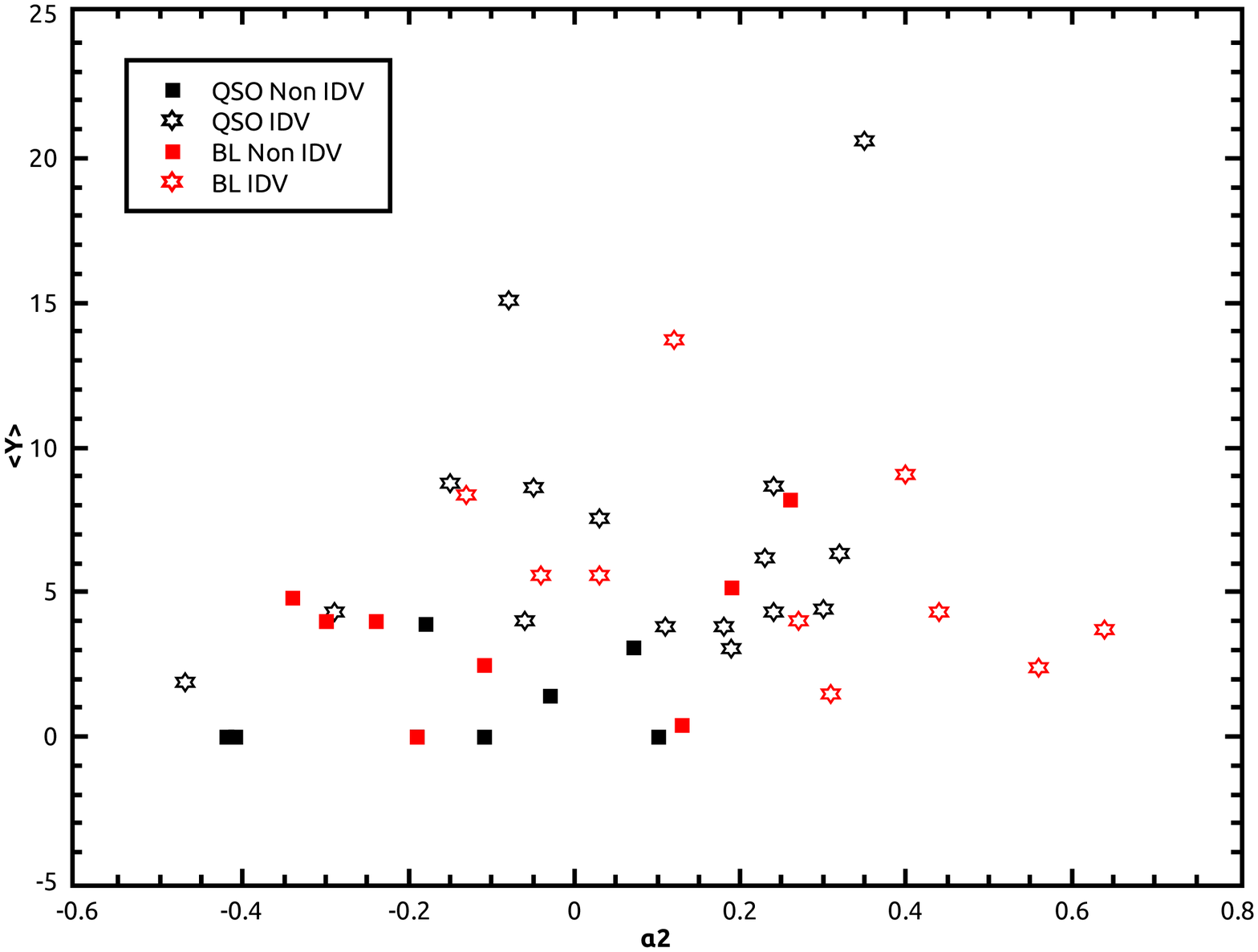}}
\caption{Intra-day variability amplitude
$\langle\,Y\,\rangle$ (averaged over all IDV sessions, except when Y = 0) versus spectral index 
($\alpha1$ from SPECFIND V2.0 and $\alpha2$ from CRATES); black: QSO, red: BL Lac; star: IDV, filled square: non-IDV.} 
\label{fig:fig3}
\end{figure}

We plot the variability strength versus the spectral index for 42
blazars in Fig.~\ref{fig:fig3}. No obvious correlation was found between the
spectral index $\alpha$ (from either the SPECFIND V2.0 catalog or
the CRATES catalog) and variability amplitude $\langle\,Y\,\rangle$ of intra-day
variability. The same results were obtained for inter-month variability (plot not shown here).
 However, it
appears that the non-IDV blazars have the spectral indices
$\alpha<0.1$ for the SPECFIND V2.0 and $\alpha<0.2$ 
(except one source with 0.26) for the CRATES 1.4/8.4 GHz spectral indices 
in Fig.~\ref{fig:fig3}, suggesting that the non-IDV blazars have
relatively `steeper' spectral indices than the IDV blazars in general.

\subsection{IDV and inter-month variability}

We tried to check whether there is a relationship between intra-day and
inter-month variability. We use only the inter-month data which
have been observed at least in 5 months. No significant correlation was found for the sub-sample of 24 sources, 
however, better results could be expected for more
sophisticated observations on a larger sample of sources in future.

\subsection{Variability strength of QSOs and BL Lacs}

We studied the different variability strength $Y$ of IDV between
QSOs and BL Lacs. Due to our incomplete sample, we calculate a
median rather than a mean of the variability index. The result
shows that the median is 3.96$\pm$4.92 for QSOs, and the median is
4.16$\pm$3.35 for BL Lacs. So there is no significant difference between
QSO and BL Lac variability strength of IDV.

All the 42 sources have been observed by VLBI techniques, e.g. in MOJAVE project 
(Monitoring Of Jets in Active galactic nuclei with VLBA Experiments, see Lister et al. 2009), 
and in the VCS project (VLBA Calibrator Survey, http://astrogeo.org/vcs/). 
We roughly checked that the VLBI structures of the 42
blazars, found that 14 out of 16 the type `c' blazars in Table~\ref{table:flux}
are IDV sources, suggesting that IDV occurs more often in the
VLBI-core dominant blazars. There is no correlation between the
IDV variability and the source redshift.

\section{Summary and outlook}

With the Urumqi telescope at 4.8\,GHz, we have measured flux
densities of 52 blazars, carried out IDV observations and
inter-month flux monitoring for 42 blazars, from the first
three-month $\it{Fermi}$-detected AGN list. We summary the results
as follows:
\begin{enumerate}
\item
26 IDV sources are detected at a high confidence level, and the
IDV detection rate of $\sim60\%$ in the $\it{Fermi}$-blazar sample is
higher than that in previous flat-spectrum AGN samples. The IDV
appears more often in the VLBI-core dominant blazars.
\item
There is a correlation between the 4.8 GHz radio flux density
and the $\gamma$-ray intensity for the 42 blazars as a whole,
in which the correlation confidence is higher for BL Lacs than that for quasars.
 \item
Pronounced inter-month variability was found in two BL Lac
objects: J0112+2244 and J0238+1636.
\item
No obvious correlation was found between the spectral indices and
modulation indices of either intra-day or inter-month variability of the blazars.
However, the non-IDV blazars tend to have relatively `steeper' spectral indices
than the IDV blazars in general.
\item
No significant correlation between the intra-day and inter-month
variability was found in the data.
\item
No significant difference was found between QSO and BL Lac variability
strength of IDV in the current sample.
\end{enumerate}

We are aware of the fact that this sample is small. Following the
pilot studies presented in this paper, we have launched a program
to search for rapid variability in a large sample of radio sources
with the Urumqi telescope in 2010, which uses the CRATES
catalog as the parent sample to further investigate our
findings. With this new project we plan to study in detail the
statistics of IDV, a possible correlation between the occurrence
of IDV and the presence of strong $\gamma$-ray emission, and the
properties of $\gamma$-ray AGN and non-$\gamma$-ray AGN.

\normalem

\begin{acknowledgements}

We thank the referee for helpful comments and suggestions. 
This work is supported by the National Natural Science Foundation
of China under grant No.11073036 and the 973 Program of China
(2009CB824800). This research has made use of data from the MOJAVE
database that is maintained by the MOJAVE team (Lister et al.,
2009, AJ, 137, 3718).

\end{acknowledgements}

\appendix


\label{lastpage}

\end{document}